\documentclass{article}

\usepackage{microtype}
\usepackage{graphicx}
\usepackage{subcaption}
\usepackage{booktabs} 
\usepackage{algorithm}
\usepackage{algorithmic}
\usepackage{newfloat}
\usepackage{listings}

\usepackage{amsfonts}

\usepackage{multirow}
\usepackage{caption}
\usepackage{lscape}
\usepackage{rotating}
\usepackage{wrapfig}

\usepackage{makecell}

\usepackage{enumitem}
\setlist[itemize]{topsep=2pt, itemsep=1pt, parsep=0pt, partopsep=0pt}

\usepackage{xcolor}

\usepackage{hyperref}

\usepackage[preprint]{icml2026}

\usepackage{amsmath}
\usepackage{amssymb}
\usepackage{mathtools}
\usepackage{amsthm}

\usepackage[capitalize,noabbrev]{cleveref}

\theoremstyle{plain}

\theoremstyle{definition}

\theoremstyle{remark}

\usepackage[textsize=tiny]{todonotes}

\begin{document}

\twocolumn[
  \icmltitle{CAFE: Channel-Autoregressive Factorized Encoding for Robust Biosignal Spatial Super-Resolution}
  \icmlsetsymbol{equal}{*}

    \begin{icmlauthorlist}
      \icmlauthor{Hongjun Liu}{equal,ustb1}
      \icmlauthor{Leyu Zhou}{equal,ustb1}
      \icmlauthor{Zijianghao Yang}{ustb1}
      \icmlauthor{Rujun Han}{ustb1}
      \icmlauthor{Shitong Duan}{ustb2}
      \icmlauthor{Kuanjian Tang}{ustb2}
      \icmlauthor{Chao Yao}{ustb2}
    \end{icmlauthorlist}
    
    \icmlaffiliation{ustb1}{School of Intelligence Science and Technology, University of Science and Technology Beijing, Beijing, China}
    \icmlaffiliation{ustb2}{School of Computer and Communication Engineering, University of Science and Technology Beijing, Beijing, China}

    \icmlcorrespondingauthor{Chao Yao}{yaochao.ustb.edu.cn}





  \icmlkeywords{Spatial Super-Resolution, EEG}

  \vskip 0.3in
]



\printAffiliationsAndNotice{\icmlEqualContribution}

\begin{abstract}
High-density biosignal recordings are critical for neural decoding and clinical monitoring, yet real-world deployments often rely on low-density (LD) montages due to hardware and operational constraints. This motivates spatial super-resolution from LD observations, but heterogeneous dependencies under sparse and noisy measurements often lead to artifact propagation and false non-local correlations.
To address this, we propose CAFE, a plug-and-play rollout generation scheme that reconstructs the full montage in geometry-aligned stages. Starting from the LD channels, CAFE first recovers nearby channels and then progressively expands to more distal regions, exploiting reliable local structure before introducing non-local interactions. During training, step-wise supervision is applied over channel groups and teacher forcing with epoch-level scheduled sampling along the group dimension is utilized to reduce exposure bias, enabling parallel computation across steps. At test time, CAFE performs an autoregressive rollout across groups, while remaining plug-and-play by reusing any temporal backbone as the shared predictor.
Evaluated on $4$ modalities and $6$ datasets, CAFE demonstrates plug-and-play generality across $3$ backbones (MLP, Conv, Transformer) and achieves consistently better reconstruction than $5$ representative baselines.
\end{abstract}

\section{Introduction}

\begin{figure}[h]
  \centering
  \includegraphics[width=\columnwidth]{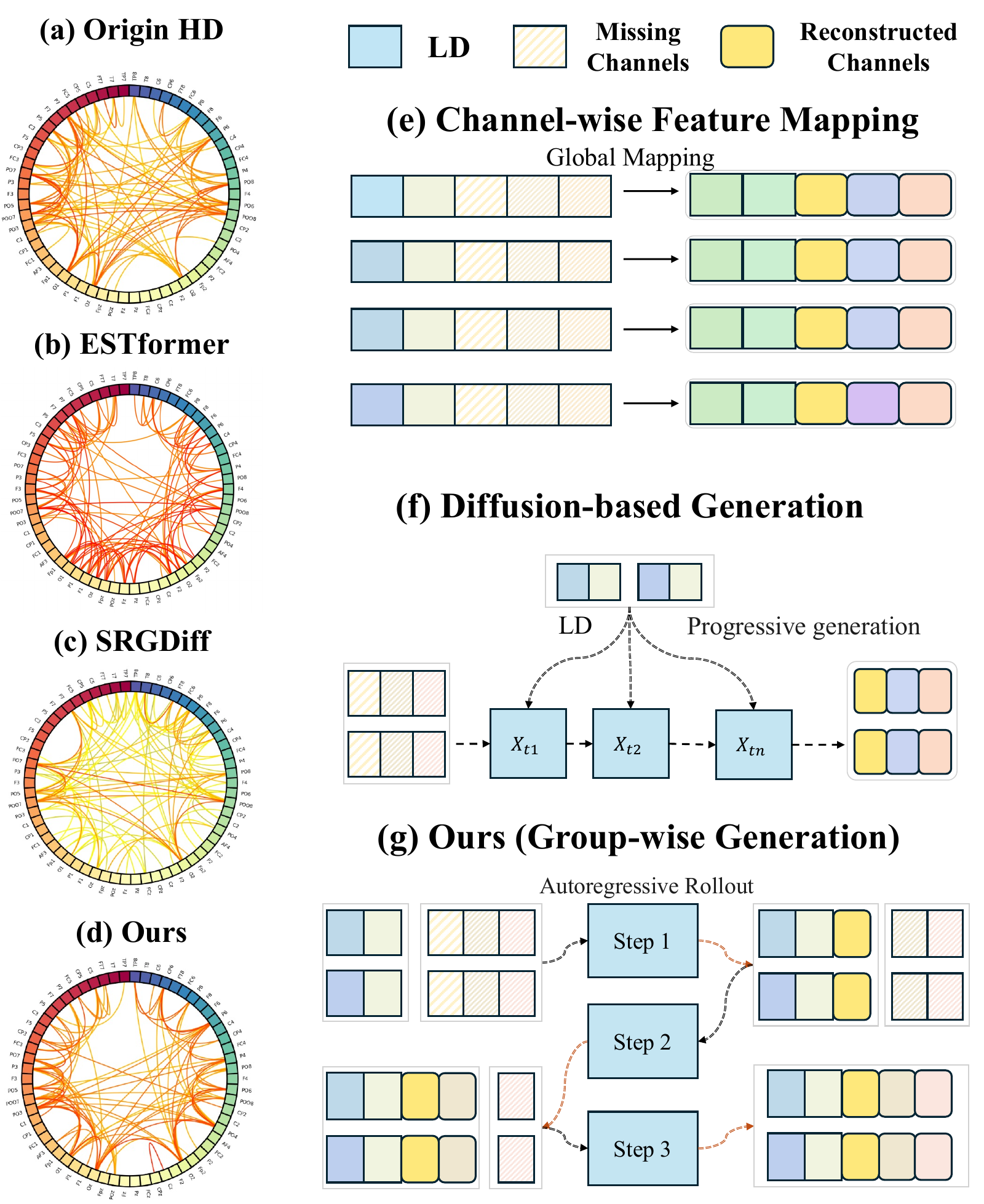}
  
  \caption{\textbf{Motivation: heterogeneous channel topology.} Top: chord diagrams illustrating that cross-channel dependencies (shown on EEG as an example) are heterogeneous across sensors, with stronger interactions concentrated within specific functional regions. Bottom: schematic comparison of channel-handling strategies. Colored blocks denote channels and their features; square blocks are features before applying a strategy, and rounded blocks are the resulting representations.}
  \vskip -0.2in
  \label{fig:difference}
\end{figure}

Multichannel biosignals and sensor arrays like electroencephalography (EEG), electromyography (EMG), electrocardiography (ECG), and electrocorticography (ECoG) could provide rich noninvasive or minimally invasive measurements of neural and physiological activity, supporting a wide range of applications in neuroscience and clinical practice, including neural decoding \citep{jiangneurolm}, clinical diagnosis \citep{liu2025spatial}, and brain–computer interfaces \citep{YIN2024110015}.
Nevertheless, their spatial resolution is inherently constrained by the number and placement of sensors and by volume‐conduction or related field–propagation effects \citep{li2025estformer}. High‐density (HD) systems with hundreds of channels can mitigate these issues but are costly, cumbersome to deploy, and uncomfortable for extended wear, whereas low‐density (LD) setups (e.g., 4 or 8 channels) are far more practical yet suffer from severe under‐sampling bias \citep{wang2025generative}. Spatial super-resolution, which reconstructs HD measurements from sparse recordings, has therefore garnered growing attention as a key tool for real-world brain–computer interfaces and broader multichannel sensing applications.

Existing work on multichannel spatial super-resolution broadly falls into three families, as depicted in Figure~\ref{fig:difference}: (i) interpolation methods \cite{dong2021reference} that estimate missing channels from spatial priors, (ii) feature mapping methods \citep{li2025estformer, 9796118} that directly regress HD signals from LD features, and (iii) generative models \citep{liu2025step, wang2025generative} including GAN- and diffusion-based models that sample HD signals conditioned on LD inputs.
Despite their architectural differences, all three families rely on a similar cross-channel dependency modeling scheme based on globally dense coupling, where channels interact via unconstrained global attention or fully-connected channel mixing, allowing information from any observed channel to directly influence any reconstructed channel. 

However, multichannel biosignals are often recorded under sparse observations and non-negligible noise due to bad leads, motion artifacts, and unstable contact~\citep{mihajlovic2014impact}, yielding incomplete and unreliable measurements across sensors. 
Moreover, cross-channel dependencies are heterogeneous: local couplings tend to be stronger and more consistent, whereas non-local interactions are weaker and less stable under sparse observations~\citep{srinivasan2007eeg} (Figure~\ref{fig:difference}(a)). 
\textbf{This motivates a reconstruction principle: rather than mixing all channels globally from the outset, the model should introduce cross-channel dependencies progressively, first exploiting reliable local structure and then incorporating more uncertain non-local interactions. }
While a key limitation of existing globally dense coupling is that it mixes all channels at once via unconstrained global attention or full channel mixing, providing no mechanism to control when and which dependencies are utilized. 

\begin{figure}[t]
  \centering
  \includegraphics[width=0.9\columnwidth]{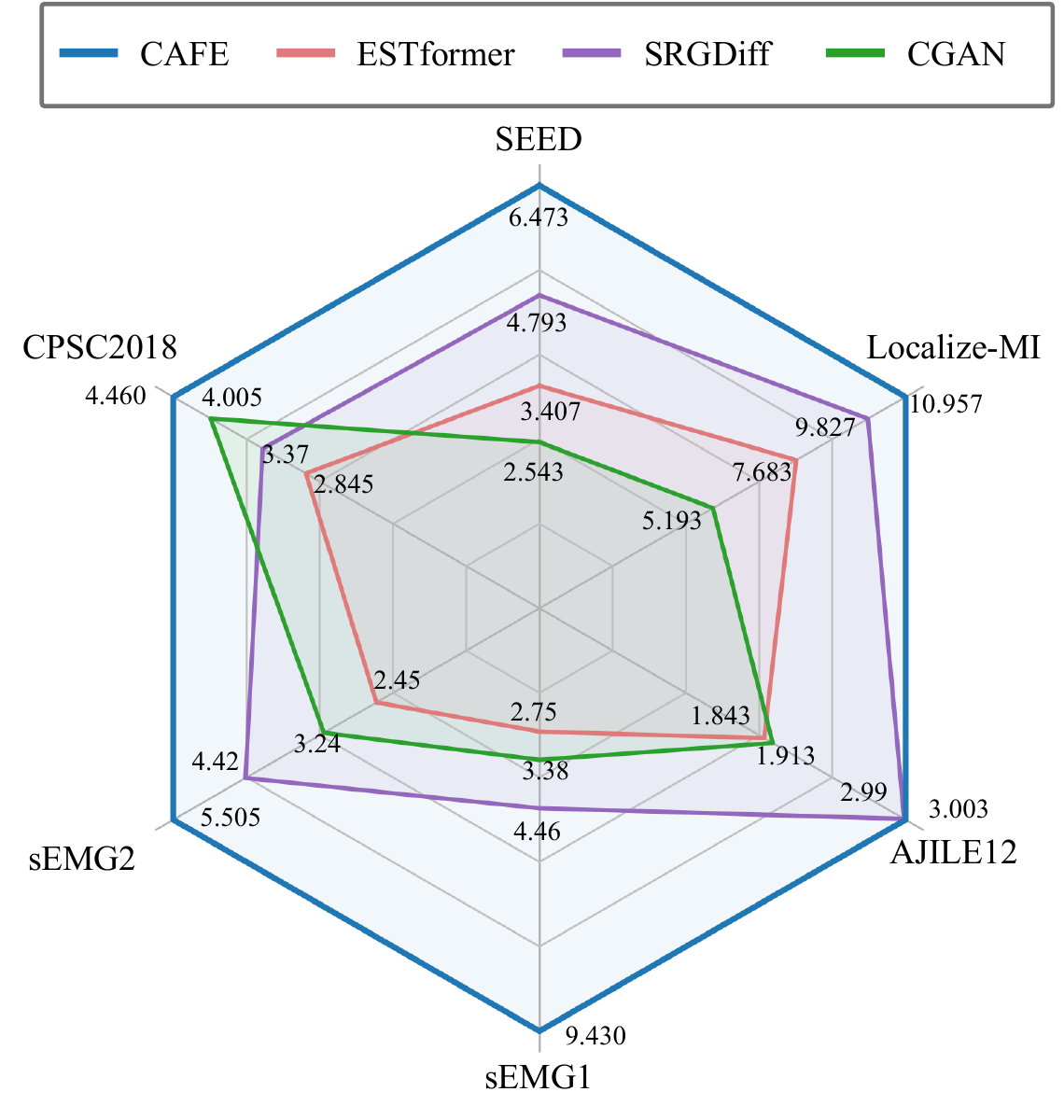}
  \caption{Performance of CAFE. Results (SNR) are averaged from all SR scale factors. CAFE outperforms strong baselines in 6 commonly used datasets.}
  \vskip -0.2in
  \label{fig:radar}
\end{figure}

To address this gap, we propose \textbf{CAFE}, a plug-and-play rollout scheme for biosignal spatial super-resolution that reconstructs missing channels in a local-to-global manner. To enforce this constraint, missing channels are partitioned into proximity-stratified, non-overlapping groups and ordered from proximal to distal with respect to the LD anchors. At each step, a backbone-agnostic shared predictor takes the visible channels as a mask-conditioned montage and outputs a full-montage estimate. During training, step-wise supervision places the loss only on the current target group, and teacher forcing is applied to construct per-step contexts, enabling parallel computation across rollout steps. To bridge the train--test gap, scheduled sampling mixes ground-truth and model-generated histories along the group dimension. At test time, an autoregressive rollout feeds previously predicted groups back into the shared predictor.
In summary, our main contributions are: 
\begin{itemize}
    \item We recast spatial super-resolution as group-wise conditional generation via a structured next-group prediction factorization.
    \item We propose a topology-aware group-wise autoregressive decoding schedule, where channels are partitioned into geometrically coherent groups using sensor distances, and groups are reconstructed in a distance-ordered sequence with a single shared predictor via multi-step refinement.
    \item We demonstrate that the group-wise autoregression rollout is backbone-agnostic by instantiating the shared predictor with representative CNN-, MLP- and Transformer-based SR models while keeping the decoding schedule unchanged.
\end{itemize}

\section{Method}

\begin{figure*}[h]
  \centering
  \includegraphics[width=2\columnwidth]{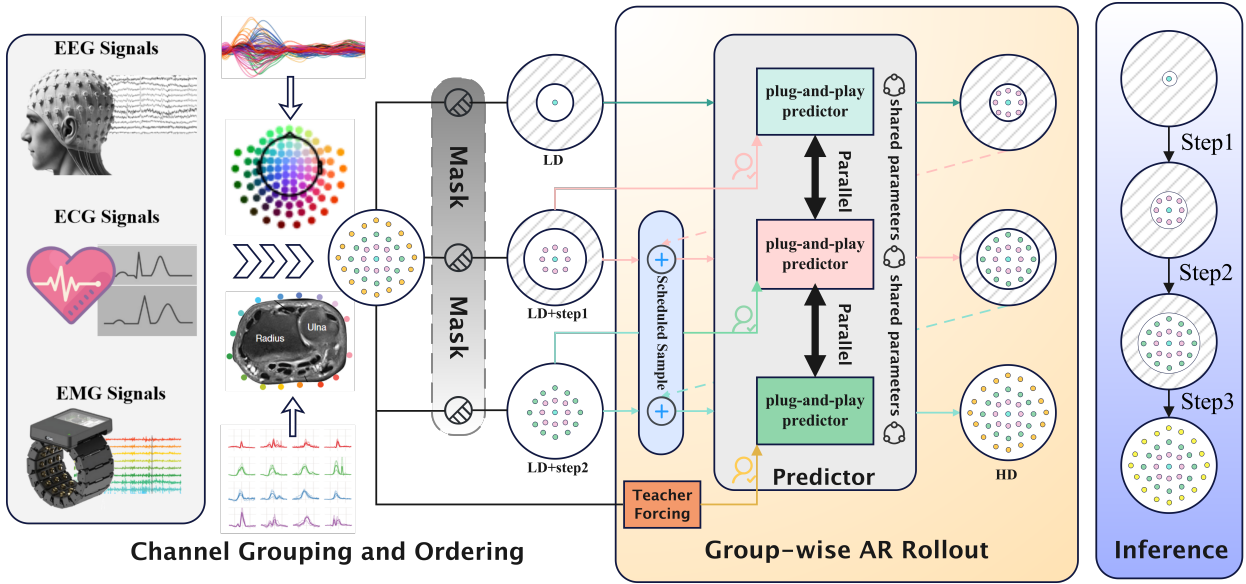}
  \caption{CAFE model overview.}
  \vskip -0.2in
  \label{fig:overview}
\end{figure*}

\subsection{Task Setup and Notation}
In this paper, biosignal spatial super-resolution is studied on a fixed canonical montage with $C_H$ channels and $T$ time samples. Let $X^H\in\mathbb{R}^{C_H\times T}$ denote the corresponding high-density (HD) signal. At test time, only a low-density (LD) subset of channels is observed. The LD channels are indexed by $\mathcal{L}$ (with $|\mathcal{L}|=C_L$) and the missing channels by $\mathcal{U}$, where $\mathcal{L}\cup\mathcal{U}$ covers the montage and $\mathcal{L}\cap\mathcal{U}=\varnothing$. Let $X_{\mathcal{L}}\in\mathbb{R}^{C_L\times T}$ denote the observed LD recordings and $X_{\mathcal{U}}\in\mathbb{R}^{(C_H-C_L)\times T}$ denote the unobserved channels. Given $X_{\mathcal{L}}$, our goal is to reconstruct the missing channels $\widehat{X}_{\mathcal{U}}$ and form the HD estimate $\widehat{X}^{H}=\operatorname{Merge}(X_{\mathcal{L}},\widehat{X}_{\mathcal{U}})$.

\subsection{Channel Grouping and Ordering}

To control which cross-channel dependencies are exposed at each prediction step and to delay potentially noisy non-local interactions, CAFE does not reconstruct all missing channels $\mathcal{U}$ at once. Instead, CAFE factorizes the reconstruction into a sequence of geometry-guided prediction steps, where each step targets a small, proximity-stratified \emph{channel group} on the canonical montage.

Given the canonical electrode coordinates ${p_i}_{i=1}^{C_H}$ and the index set $\mathcal{L}\subseteq{1,\dots,C_H}$ of observed LD channels, CAFE groups the missing channels by their Euclidean distance to the observed channels. For each missing channel $u\in\mathcal{U}$, average distance to the observed channels is computed:
\begin{equation}
d(u) \;=\; \frac{1}{|\mathcal{L}|}\sum_{\ell\in\mathcal{L}} \|p_u - p_\ell\|_2.
\label{eq:channel-to-anchor-distance}
\end{equation}

CAFE then sorts the missing-channel set $\mathcal U$ by ascending $d(u)$ and partition the ordered list into $G$ contiguous groups, where $G$ denotes the rollout depth, i.e., the number of prediction steps. Let the ordered missing-channel list be $\big(u_{(1)},\ldots,u_{(N)}\big)$ with $N=|\mathcal U|$ such that
$d\!\left(u_{(1)}\right)\le \cdots \le d\!\left(u_{(N)}\right)$.
To define the partition, CAFE chooses fixed split fractions $0<\beta_1<\cdots<\beta_{G-1}<1$ and sets boundary indices
$b_g=\lfloor \beta_g N\rfloor$ for $g=1,\ldots,G-1$, with $b_0=0$ and $b_G=N$.
The $g$-th target group is then the consecutive block
\begin{equation}
\mathcal U_g=\{u_{(i)}:\ b_{g-1}< i \le b_g\},
\label{eq:group-partition}
\end{equation}
which yields a disjoint partition $\mathcal U=\biguplus_{g=1}^{G}\mathcal U_g$.
In our experiments, a small fixed rollout depth $G=3$ and fixed split fractions $\beta_1=\frac{1}{6}$ and $\beta_2=\frac{1}{2}$ across datasets are utilized.

\subsection{Group-wise Autoregression Rollout}
\label{sec:rollout}

Given the ordered channel groups $\{\mathcal{U}_g\}_{g=1}^G$, CAFE reconstructs the full montage via a $G$-step rollout using a single shared predictor $f_\theta$ reused across steps. The autoregression is along the channel-group dimension rather than the time axis. The key idea is to maintain a step-wise visible set that grows monotonically: at step $g$, the predictor can condition only on the observed LD channels and the groups reconstructed in earlier steps, while all future groups remain hidden. Formally, the step-$g$ visible index set is defined as
\begin{equation}
\mathcal{V}_g \;=\; \mathcal{L}\ \cup\ \bigcup_{k<g}\mathcal{U}_k,
\label{eq:visible-set}
\end{equation}
and encode it by a binary mask $M_g\in\{0,1\}^{C_H}$ on the canonical montage, where $(M_g)_i=\mathbb{I}[i\in\mathcal{V}_g]$.

To keep a stable channel-aligned feature space as $\mathcal{V}_g$ expands, CAFE operates on a canonical montage and maintain an assembled full-channel tensor $\widetilde{X}^{(g)}\in\mathbb{R}^{C_H\times T}$. The tensor is initialized by placing the observed LD channels into canonical channel order and zero-filling the remaining channels:
\begin{equation}
\widetilde{X}^{(0)}[i]=
\begin{cases}
X_{\mathcal{L}}[i], & i\in\mathcal{L},\\
0, & i\in\mathcal{U}.
\end{cases}
\label{eq:rollout-init}
\end{equation}
At step $g$, a masked full-montage input is formed as $C_g \;=\; \left(M_g \mathbf{1}_T^\top\right)\odot \widetilde{X}^{(g-1)} \in \mathbb{R}^{C_H\times T}.
$, and the shared predictor produces a full-montage estimate $\widehat{X}^{(g)} = f_\theta(C_g)$. The assembled montage is then updated by writing the current prediction only to the target group while leaving all other channels unchanged:
\begin{equation}
\widetilde{X}^{(g)}[i]=
\begin{cases}
X_{\mathcal{L}}[i], & i\in\mathcal{L},\\
\widehat{X}^{(g)}[i], & i\in\mathcal{U}_g,\\
\widetilde{X}^{(g-1)}[i], & \text{otherwise}.
\end{cases}
\label{eq:rollout-merge}
\end{equation}
After $G$ steps, the final HD estimate is $\widehat{X}^{H}=\widetilde{X}^{(G)}$.
\begin{equation}
\widehat{X}^{H}=\operatorname{Merge}(X_{\mathcal{L}},\widehat{X}_{\mathcal{U}})
=\widetilde{X}^{(G)}.
\label{eq:final-merge-rollout}
\end{equation}

The same rollout is utilized for training and inference, and the only difference is how the history on $\mathcal{U}_{<g}$ is provided when forming $C_g$. During training, teacher forcing and scheduled sampling schemes are utilized to construct $\widetilde{X}^{(g-1)}$ from ground truth and model predictions. During inference, the rollout is fully self-conditioned and $\widetilde{X}^{(g-1)}$ contains only the anchors and the predictions from earlier steps. Since the rollout schedule, masking rule, and step-wise supervision are defined outside $f_\theta$, any temporal backbone that maps a mask-conditioned canonical montage to a full-montage output can serve as $f_\theta$, making the scheme plug-and-play. 
Appendix further formalizes the rollout and shows how conditional uncertainty decreases as the visible set expands.

\begin{table*}[t]
\centering
\small
\setlength{\tabcolsep}{2.6pt}
\caption{CAFE improves single-pass reconstruction across architectures via a channel-group autoregressive (AR) rollout. ``Orig'' is the one-shot full-montage output, ``+AR'' applies the CAFE rollout, and ``Gain'' is the relative change (\%) from Orig to +AR.}
\begin{tabular}{ll ccc ccc ccc ccc ccc ccc}
\toprule
\multirow{2}{*}{Model} & \multirow{2}{*}{Metric}
& \multicolumn{3}{c}{SEED}
& \multicolumn{3}{c}{Localize-MI}
& \multicolumn{3}{c}{AJILE12}
& \multicolumn{3}{c}{sEMG1}
& \multicolumn{3}{c}{sEMG2}
& \multicolumn{3}{c}{CPSC2018} \\
\cmidrule(lr){3-5}\cmidrule(lr){6-8}\cmidrule(lr){9-11}\cmidrule(lr){12-14}\cmidrule(lr){15-17}\cmidrule(lr){18-20}
& 
& Orig & +AR & Gain$\uparrow$
& Orig & +AR & Gain$\uparrow$
& Orig & +AR & Gain$\uparrow$
& Orig & +AR & Gain$\uparrow$
& Orig & +AR & Gain$\uparrow$
& Orig & +AR & Gain$\uparrow$ \\
\midrule

\multirow{3}{*}{Conv}
& NMSE$\downarrow$
& 0.18 & 0.12 & 33\%
& 0.024 & 0.016 & 33\%
& 0.34 & 0.33 & 3\%
& 0.20 & 0.17 & 15\%
& 0.10 & 0.08 & 20\%
& 0.41 & 0.41 & 0\% \\
& PCC$\uparrow$
& 0.86 & 0.93 & 8\%
& 0.96 & 0.98 & 2\%
& 0.80 & 0.83 & 4\%
& 0.91 & 0.92 & 1\%
& 0.96 & 0.97 & 1\%
& 0.69 & 0.71 & 3\% \\
& SNR$\uparrow$
& 6.98 & 9.04 & 30\%
& 16.12 & 17.91 & 11\%
& 4.66 & 5.01 & 8\%
& 6.90 & 7.63 & 11\%
& 9.94 & 10.81 & 9\%
& 3.76 & 3.87 & 3\% \\
\cmidrule(lr){1-20}

\multirow{3}{*}{MLP}
& NMSE$\downarrow$
& 0.24 & 0.14 & 42\%
& 0.042 & 0.024 & 43\%
& 0.47 & 0.35 & 26\%
& 0.35 & 0.35 & 0\%
& 0.14 & 0.09 & 36\%
& 0.49 & 0.47 & 4\% \\
& PCC$\uparrow$
& 0.85 & 0.88 & 4\%
& 0.96 & 0.96 & 0\%
& 0.73 & 0.81 & 11\%
& 0.81 & 0.81 & 0\%
& 0.94 & 0.96 & 2\%
& 0.60 & 0.61 & 2\% \\
& SNR$\uparrow$
& 6.12 & 8.19 & 34\%
& 13.73 & 16.17 & 18\%
& 3.26 & 4.94 & 52\%
& 4.54 & 4.59 & 1\%
& 8.47 & 10.23 & 21\%
& 3.09 & 3.24 & 5\% \\
\cmidrule(lr){1-20}

\multirow{3}{*}{Trans}
& NMSE$\downarrow$
& 0.22 & 0.14 & 36\%
& 0.034 & 0.022 & 35\%
& 0.45 & 0.35 & 22\%
& 0.35 & 0.33 & 6\%
& 0.13 & 0.08 & 38\%
& 0.48 & 0.47 & 2\% \\
& PCC$\uparrow$
& 0.86 & 0.88 & 2\%
& 0.97 & 0.97 & 0\%
& 0.74 & 0.81 & 9\%
& 0.81 & 0.82 & 1\%
& 0.94 & 0.97 & 3\%
& 0.61 & 0.62 & 2\% \\
& SNR$\uparrow$
& 6.50 & 8.24 & 27\%
& 14.63 & 16.56 & 13\%
& 3.43 & 4.90 & 43\%
& 4.51 & 4.75 & 5\%
& 8.71 & 10.73 & 23\%
& 3.10 & 3.27 & 5\% \\
\bottomrule
\end{tabular}
\label{tab:arch_promotion}
\end{table*}

\subsection{Training with teacher forcing and scheduled sampling}
\label{sec:training}

\paragraph{Objective.}
Training supervises only the current target group at each step, yielding a next-group reconstruction objective under the group-wise autoregressive rollout with a shared predictor. The step-$g$ context $C_g $ is obtained by masking the assembled montage with the availability mask where $M_g\in\{0,1\}^{C_H}$ is broadcast along the time axis to match the $C_H\times T$ shape, and $\odot$ denotes element-wise multiplication. The objective is
\begin{equation}
\min_{\theta}\;
\mathbb{E}_{(X_{\mathcal L},X^H)\sim\mathcal D}
\Bigg[
\sum_{g=1}^{G}
\big\|
f_\theta(C_g)[\mathcal{U}_g]-X_{\mathcal U_g}
\big\|_2^2
\Bigg].
\label{eq:train_obj}
\end{equation}

\paragraph{Teacher forcing.}
To train each step under oracle history, the context is constructed with ground-truth previous groups $X_{\mathcal U_{<g}}$:
\begin{equation}
\widehat{X}^{\text{TF}(g)} = f_\theta(C_g^{\text{TF}}),
\qquad
\widehat{X}_{\mathcal U_g}^{\text{TF}} = \widehat{X}^{\text{TF}(g)}[\mathcal U_g].
\label{eq:tf}
\end{equation}
This trains $f_\theta$ under clean contexts and yields stable conditional mappings for each group.

\paragraph{Scheduled sampling.}
At test time, $C_g$ must be built from model predictions, which can introduce exposure bias.
To approximate this condition while keeping training parallelizable across groups, we adopt an
epoch-level scheduled sampling strategy along the group dimension.
Specifically, we maintain a per-sample cache of predicted missing groups from the previous epoch,
denoted as $\bar X_{\mathcal U_k}^{(e-1)}$, obtained by running one autoregressive rollout with the
model snapshot $\theta^{(e-1)}$ (i.e., using only predicted previous groups, $\pi=0$).

At epoch $e$, for each $k<g$ we sample $z_{e,k}\sim\mathrm{Bernoulli}(\pi)$ and form a mixed history:
\begin{equation}
\widetilde X_{\mathcal U_k}^{(e)}
= z_{e,k}\, X_{\mathcal U_k} + (1-z_{e,k})\, \bar X_{\mathcal U_k}^{(e-1)}.
\label{eq:ss_mix_cache}
\end{equation}
Using $\{\widetilde X_{\mathcal U_k}^{(e)}\}_{k<g}$ to build $C_g^{(e)}$, we predict
\begin{equation}
\widehat{X}^{(e,g)} = f_\theta(C_g^{(e)}),\qquad
\widehat{X}_{\mathcal U_g}^{(e,g)} = \widehat{X}^{(e,g)}[\mathcal U_g].
\label{eq:ss_pred_cache}
\end{equation}
Since the cache $\bar X^{(e-1)}$ is fixed with respect to the current parameters $\theta$,
all contexts $\{C_g^{(e)}\}_{g=1}^G$ can be constructed before the forward pass, enabling parallel
computation across groups by stacking $\{C_g^{(e)}\}$ along the batch dimension.
After finishing epoch $e$, we refresh the cache by performing one rollout with $\pi=0$ using $\theta^{(e)}$,
yielding $\bar X^{(e)}$ for the next epoch.
We use a constant sampling rate $\pi=0.95$ during training.
At inference, $\pi=0$ and only predicted previous groups are used.

\section{Experiments}
\subsection{Experiment Setup}

\paragraph{Datasets.}
To assess the effectiveness of the proposed CAFE, we evaluate our method on $6$ publicly available multichannel biosignal datasets: two high-density EEG benchmarks (\textbf{SEED}, \textbf{Localize-MI}); two high-density \textbf{sEMG} datasets for discrete gestures and handwriting \citep{kaifosh2025generic}; the \textbf{AJILE12} intracranial ECoG corpus \citep{peterson2022ajile12}; and a 12-lead ECG dataset (\textbf{CPSC2018}). Detailed dataset descriptions and preprocessing protocols are provided in Appendix.

\paragraph{Experiment Settings.}
For each dataset, multiple upsampling factors are employed (SEED: $2/4/8\times$; Localize-MI: $2/4/8/16\times$; AJILE12: $2/4/8\times$; sEMG: $2/4\times$; CPSC2018: $4/12\times$) by selecting $C_{\mathrm{HD}}/s$ observed channels per factor and treating the rest as missing. We follow the evaluation protocol of SRGDiff and ESTformer and use four predefined LD channel layouts that maximize spatial coverage. All methods are evaluated on the same four layouts for fair comparison. Unless otherwise stated, averaged results over the four layouts and test subjects are reported (Detailed settings in Appendix). The hyperparameter settings of CAFE are summarized in Appendix.

\paragraph{Metrics.}
CAFE is evaluated at three complementary levels of SR quality. \textbf{Signal fidelity:} NMSE, PCC, and reconstruction SNR w.r.t.\ the HD reference. \textbf{Feature \& spectral fidelity:} EEG-FID computed using a frozen EEGNet trained on each dataset's training split, and a frequency-domain distortion metric based on STFT power spectra (Spec-MAE). \textbf{Downstream utility:} accuracy on SEED subject-dependent emotion recognition and binary epileptic classification on Localize-MI. Unless otherwise stated, the mean results over subjects and LD layouts are reported in the main tables. Metric definitions and implementation details are provided in Appendix.

\subsection{Autoregressive Rollout Comparisons}
\label{sec:autoregression_analysis}

Given the relative novelty of group-wise autoregressive generation for multichannel spatial super-resolution, a systematic analysis of the factors underlying its effectiveness is conducted. Specifically, we examine
(i) generalization across representative backbone families,
(ii) the role of reconstruction ordering,
(iii) mitigation of exposure bias via different rollout scheme, and
(iv) the trade-off induced by channel-group size (i.e., autoregressive step length).

\subsubsection{Backbone Generalization}
Our proposed group-wise autoregressive rollout is incorporated into three representative architectures commonly used for spatial super-resolution: (i) a depthwise-separable convolutional encoder (\textbf{Conv}), (ii) a lightweight MLP with channel-wise positional encodings (\textbf{MLP}), and (iii) a standard time series transformer (\textbf{Transformer}). This experiment isolates whether the performance gains stem from the structured decoding factorization itself rather than backbone-specific inductive biases. For each backbone, the performance of the original model, its autoregressive-augmented counterpart, and the relative improvement (\textit{Gain}) on NMSE, PCC, and SNR are reported. To make the improvement more apparent, \textit{Gain} is computed as $(m_{\mathrm{orig}}-m_{\mathrm{AR}})/m_{\mathrm{orig}}$ for metrics where lower is better. Table~\ref{tab:arch_promotion} shows that autoregressive rollout consistently improves reconstruction quality across all three backbone families, with the largest relative gains observed on high-density modalities, indicating that structured dependency expansion is particularly beneficial when rich spatial structure is present. 
Smaller gains are observed on CPSC2018, likely because ECG leads are defined as voltage differences between electrode pairs, yielding weaker and less consistent spatial coupling across channels.

\begin{figure}[h]
  \centering
  \includegraphics[width=\columnwidth]{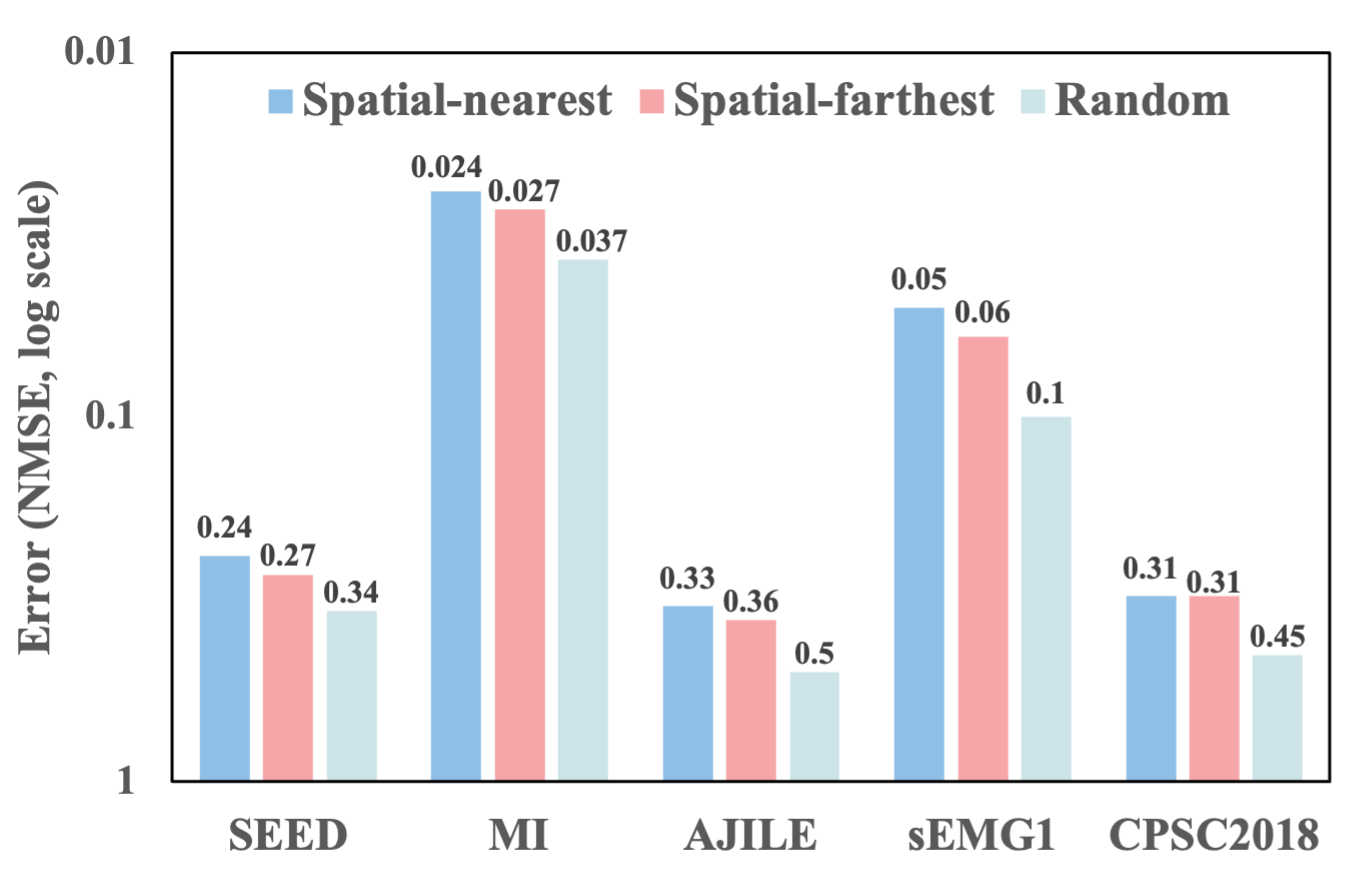}
  \caption{Effect of decoding order in group-wise autoregressive reconstruction.}
  \label{fig:ordering}
\end{figure}

\subsubsection{Reconstruction Ordering}
Our method claims that the gain comes from topology-aware progressive conditioning rather than autoregression alone. Therefore, if ordering is a causal factor, reversing or randomizing the group order should noticeably degrade performance; otherwise, the results should be largely order-invariant.
Thus, ordering sensitivity is evaluated by comparing: (i) a \emph{proximal-to-distal} schedule that reconstructs groups from nearest to farthest, (ii) a \emph{distal-to-proximal} reversed schedule, and (iii) a \emph{random} schedule with uniformly shuffled group order. Figure~\ref{fig:ordering} shows that the proximal-to-distal ordering consistently yields the lowest NMSE, while distal-to-proximal and random schedules degrade performance. This supports our hypothesis that prematurely exposing the predictor to unreliable long-range dependencies amplifies error propagation under sparse observations; in contrast, expanding context from reliable local groups provides better-conditioned inputs for subsequent distant groups. Complementary cumulative-error evidence across datasets are provided in Appendix.

\begin{figure}[h]
  \centering
  \includegraphics[width=\columnwidth]{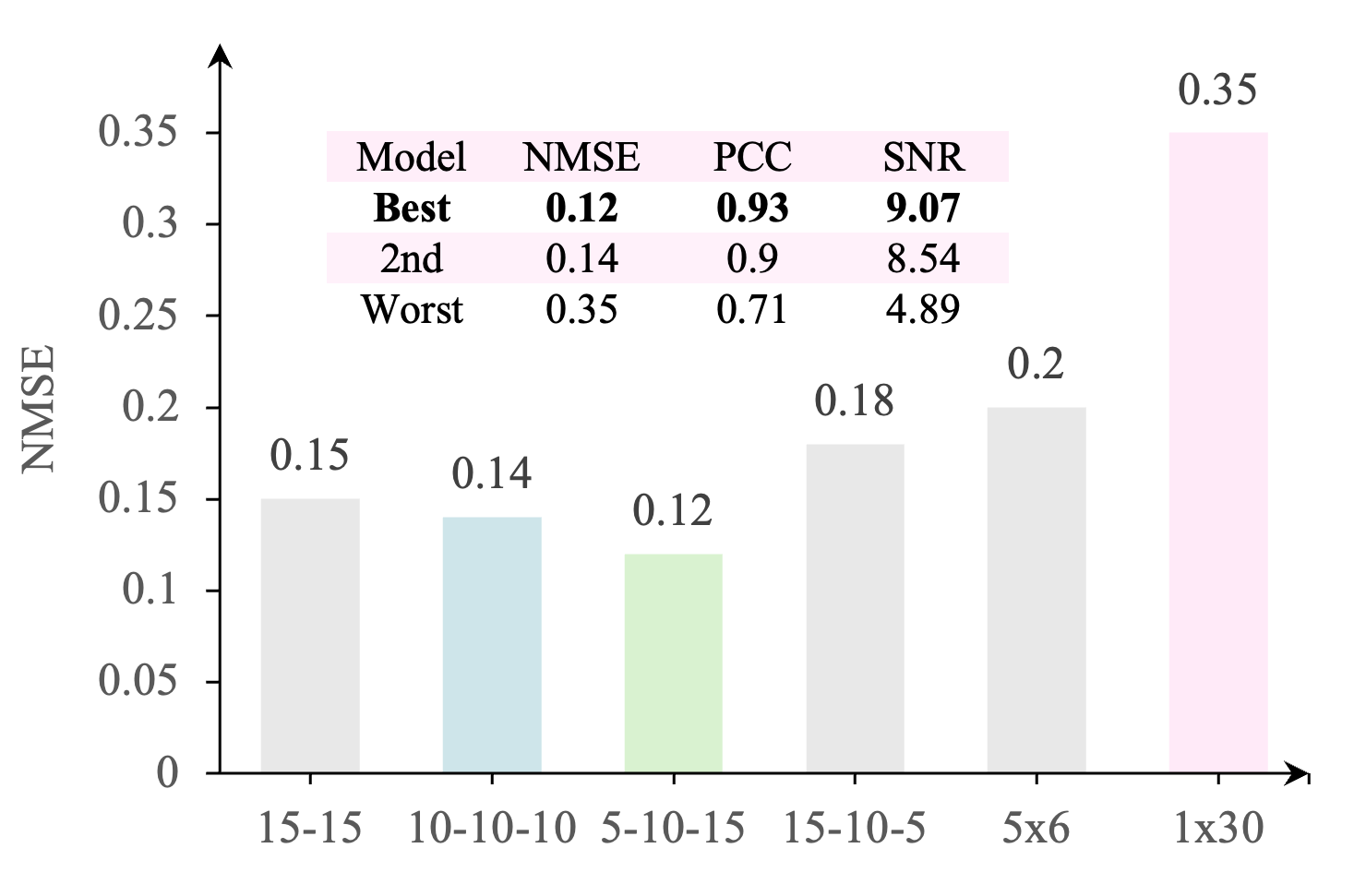}
  \vskip -0.2in
  \caption{Effect of rollout granularity (channels per step) on SEED.}
  \vskip -0.1in
  \label{fig:granularity}
\end{figure}

\begin{table*}[h]
\centering
\small
\setlength{\tabcolsep}{3pt}
\caption{Cross-subject channel super-resolution results (NMSE/PCC/SNR) on $6$ multichannel biosignal datasets.}
\begin{tabular}{ll ccc cccc ccc cc cc cc}
\toprule
\multicolumn{2}{c}{} 
& \multicolumn{3}{c}{SEED}
& \multicolumn{4}{c}{Localize-MI}
& \multicolumn{3}{c}{AJILE12}
& \multicolumn{2}{c}{sEMG1}
& \multicolumn{2}{c}{sEMG2}
& \multicolumn{2}{c}{CPSC2018} \\
\cmidrule(lr){3-5}
\cmidrule(lr){6-9}
\cmidrule(lr){10-12}
\cmidrule(lr){13-14}
\cmidrule(lr){15-16}
\cmidrule(lr){17-18}
Method & Metric
& 2$\times$ & 4$\times$ & 8$\times$
& 2$\times$ & 4$\times$ & 8$\times$ & 16$\times$
& 2$\times$ & 4$\times$ & 8$\times$
& 2$\times$ & 4$\times$
& 2$\times$ & 4$\times$
& 4$\times$ & 12$\times$ \\
\midrule

\multirow{3}{*}{GRIN}
& NMSE
& 0.40 & 0.43 & 0.54
& 0.07 & 0.13 & 0.19 & 0.27
& 0.56 & 0.81 & 0.92
& 0.55 & 0.75
& 0.43 & 0.65
& 0.53 & 0.79 \\
& PCC
& 0.76 & 0.72 & 0.66
& 0.93 & 0.91 & 0.89 & 0.80
& 0.65 & 0.56 & 0.23
& 0.67 & 0.55
& 0.75 & 0.60
& 0.68 & 0.57 \\
& SNR
& 4.06 & 3.58 & 2.79
& 12.13 & 8.57 & 6.82 & 5.53
& 2.57 & 0.89 & 0.43
& 2.63 & 1.92
& 3.39 & 1.76
& 2.72 & 1.16 \\
\cmidrule(lr){1-18}

\multirow{3}{*}{CGAN}
& NMSE
& 0.48 & 0.53 & 0.67
& 0.14 & 0.28 & 0.41 & 0.49
& 0.43 & 0.75 & 0.82
& 0.38 & 0.55
& 0.38 & 0.58
& \underline{0.36} & \underline{0.43} \\
& PCC
& 0.73 & 0.70 & 0.65
& 0.88 & 0.79 & 0.77 & 0.72
& 0.73 & 0.51 & 0.40
& 0.79 & 0.67
& 0.79 & 0.65
& \underline{0.78} & \underline{0.73} \\
& SNR
& 3.16 & 2.74 & 1.73
& 8.41 & 5.46 & 3.83 & 3.07
& 3.64 & 1.24 & 0.86
& 4.18 & 2.58
& 4.15 & 2.33
& \underline{4.39} & \underline{3.62} \\
\cmidrule(lr){1-18}

\multirow{3}{*}{TimeMixer++}
& NMSE
& 0.55 & 0.64 & 0.72
& 0.17 & 0.29 & 0.41 & 0.53
& 0.76 & 0.89 & 0.95
& 0.61 & 0.83
& 0.68 & 0.87
& 0.66 & 0.85 \\
& PCC
& 0.64 & 0.57 & 0.51
& 0.91 & 0.84 & 0.77 & 0.66
& 0.49 & 0.32 & 0.21
& 0.65 & 0.39
& 0.57 & 0.33
& 0.56 & 0.36 \\
& SNR
& 2.62 & 2.05 & 1.49
& 8.16 & 5.37 & 3.76 & 2.41
& 1.17 & 0.58 & 0.31
& 2.31 & 0.86
& 1.52 & 0.64
& 1.77 & 0.79 \\
\cmidrule(lr){1-18}

\multirow{3}{*}{ESTformer}
& NMSE
& 0.44 & 0.47 & 0.56
& 0.09 & 0.16 & 0.22 & 0.28
& 0.46 & 0.72 & 0.85
& 0.44 & 0.63
& 0.47 & 0.68
& 0.44 & 0.61 \\
& PCC
& 0.74 & 0.72 & 0.65
& 0.95 & 0.92 & 0.88 & 0.85
& 0.74 & 0.53 & 0.38
& 0.75 & 0.62
& 0.73 & 0.58
& 0.75 & 0.63 \\
& SNR
& 3.95 & 3.54 & 2.73
& 10.44 & 8.08 & 6.58 & 5.63
& 3.40 & 1.43 & 0.70
& 3.52 & 1.98
& 3.25 & 1.65
& 3.55 & 2.14 \\
\cmidrule(lr){1-18}

\multirow{3}{*}{SRGdiff}
& NMSE
& \underline{0.27} & \underline{0.38} & \underline{0.45}
& \underline{0.04} & \underline{0.06} & \underline{0.19} & \underline{0.23}
& \underline{0.36} & \underline{0.65} & \underline{0.73}
& \underline{0.26} & \underline{0.49}
& \underline{0.27} & \underline{0.48}
& 0.40 & 0.52 \\
& PCC
& \underline{0.80} & \underline{0.72} & \underline{0.69}
& \underline{0.96} & \underline{0.93} & \underline{0.90} & \underline{0.86}
& \underline{0.80} & \underline{0.59} & \underline{0.52}
& \underline{0.87} & \underline{0.71}
& \underline{0.86} & \underline{0.71}
& 0.75 & 0.67 \\
& SNR
& \underline{5.69} & \underline{4.57} & \underline{4.12}
& \underline{13.65} & \underline{12.15} & \underline{7.21} & \underline{6.30}
& \underline{4.89} & \underline{1.97} & \underline{1.39}
& \underline{5.83} & \underline{3.09}
& \underline{5.67} & \underline{3.17}
& 3.93 & 2.81 \\
\cmidrule(lr){1-18}

\multirow{3}{*}{CAFE}
& NMSE
& \textbf{0.12} & \textbf{0.24} & \textbf{0.37}
& \textbf{0.02} & \textbf{0.04} & \textbf{0.14} & \textbf{0.20}
& \textbf{0.33} & \textbf{0.59} & \textbf{0.67}
& \textbf{0.05} & \textbf{0.23}
& \textbf{0.19} & \textbf{0.41}
& \textbf{0.31} & \textbf{0.41} \\
& PCC
& \textbf{0.91} & \textbf{0.87} & \textbf{0.75}
& \textbf{0.98} & \textbf{0.94} & \textbf{0.91} & \textbf{0.89}
& \textbf{0.83} & \textbf{0.63} & \textbf{0.56}
& \textbf{0.96} & \textbf{0.87}
& \textbf{0.89} & \textbf{0.76}
& \textbf{0.83} & \textbf{0.76} \\
& SNR
& \textbf{9.07} & \textbf{6.15} & \textbf{4.20}
& \textbf{15.32} & \textbf{13.18} & \textbf{8.54} & \textbf{6.79}
& \textbf{5.01} & \textbf{2.10} & \textbf{1.90}
& \textbf{12.56} & \textbf{6.30}
& \textbf{7.20} & \textbf{3.81}
& \textbf{5.06} & \textbf{3.86} \\
\bottomrule
\end{tabular}
\label{tab:cross_main}
\end{table*}

\begin{table*}[t]
\centering
\small
\caption{{Frequency-domain MAE between reconstructed and real HD topomaps on SEED, Localize-MI and AJILE12.}}
\label{tab:topomap}
\setlength{\tabcolsep}{4pt}
\begin{tabular}{l lcccccccccc}
\toprule
\multirow{2}{*}{Model} &
\multirow{2}{*}{Reference} &
\multicolumn{3}{c}{SEED} &
\multicolumn{4}{c}{Localize-MI ($\times 10^{-7}$)} &
\multicolumn{3}{c}{AJILE12} \\
\cmidrule(lr){3-5} \cmidrule(lr){6-9} \cmidrule(lr){10-12}
 &  & 2$\times$ & 4$\times$ & 8$\times$
 & 2$\times$ & 4$\times$ & 8$\times$ & 16$\times$
 & 2$\times$ & 4$\times$ & 8$\times$ \\
\midrule
GRIN
 & ICLR 2022
 & 0.827  & 0.856  & 0.883
 & 0.347  & 0.615  & 0.832 & 0.953
 & 0.572  & 0.748  & 0.983 \\
CGAN
 & NeurIPS 2024
 & 0.474  & 0.497  & 0.518
 & 0.113  & 0.517  & 0.633 & 0.763
 & 0.287  & 0.519  & 0.746 \\
TimeMixer++
 & ICLR 2025
 & 1.372  & 1.528  & 1.753
 & 0.434  & 0.976  & 1.361 & 1.832
 & 0.756  & 0.947  & 1.462 \\
ESTformer
 & KBS 2025
 & 0.511  & 0.520  & 0.613
 & 0.145  & 0.550  & 0.777 & 0.823
 & 0.392  & 0.589  & 0.849 \\
SRGdiff
 & ICLR 2026
 & \underline{0.376}  & \underline{0.433}  & \underline{0.452}
 & \underline{0.108}  & \underline{0.413}  & \underline{0.592} & \underline{0.716}
 & \underline{0.257}  & \underline{0.431}  & \underline{0.529} \\
\textbf{CAFE}
 & Ours
 & \textbf{0.28}  & \textbf{0.35}  & \textbf{0.266}
 & \textbf{0.028}  & \textbf{0.074}  & \textbf{0.087} & \textbf{0.143}
 & \textbf{0.186}  & \textbf{0.378} & \textbf{0.462} \\
\bottomrule
\end{tabular}
\end{table*}

\subsubsection{Rollout Granularity}
CAFE uses multi-step decoding where earlier predicted groups become part of the context for later steps. Hence, the step size controls a fundamental trade-off: smaller groups reduce per-step uncertainty but increase rollout length and error accumulation, while larger groups shorten the chain but weaken the benefit of progressive conditioning. Therefore, the number of channels generated per step is varied while keeping the total number of reconstructed channels fixed. As shown in Figure~\ref{fig:granularity}, moderate three-step schedules ($5$--$10$--$15$) achieve the best NMSE. Extremely fine-grained decoding with $1\times30$ performs worse because it requires a 30-step rollout that predicts one channel per step, making the reconstruction highly sensitive to accumulated errors and exposure bias. Rollouts such as $5\times6$ also degrade relative to $3$ steps, as the longer decoding chain still compounds intermediate errors. The same trend holds on additional datasets, with results reported in Appendix.

\subsubsection{Rollout Scheme}
\begin{figure*}[h]
  \centering
  \includegraphics[width=2\columnwidth]{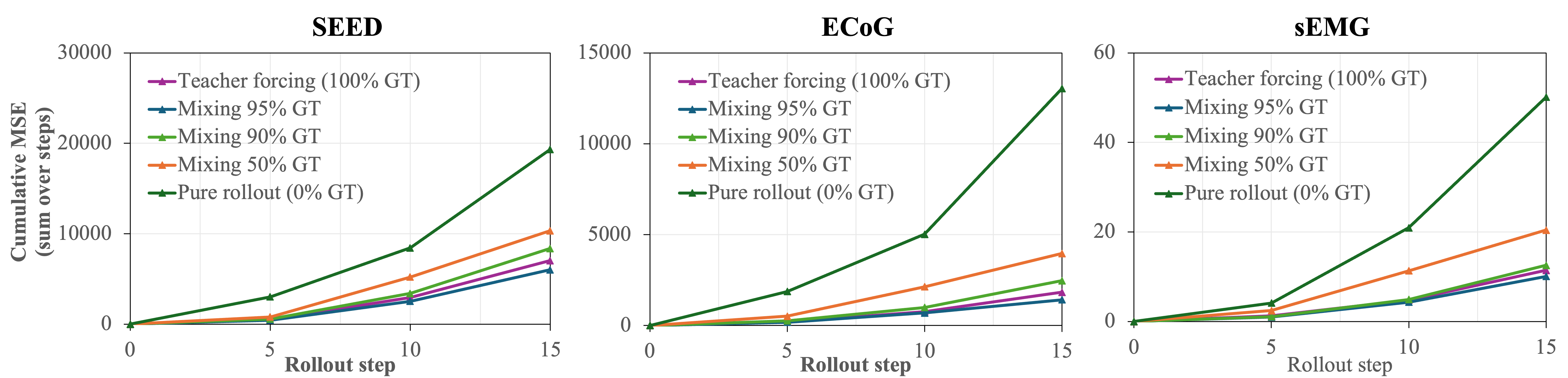}
  \caption{Training-context ablation for group-wise rollouts on SEED, ECoG, and sEMG. We plot cumulative NMSE up to rollout step $g$ ($\sum_{i=1}^{g}\mathrm{NMSE}_i$) under teacher forcing (100\% GT), scheduled sampling (mixed GT), and pure rollout (0\% GT). Lower is better.}
  \label{fig:sample}
\end{figure*}

Because CAFE performs multi-step decoding, the training context (ground-truth vs. predicted history) directly determines the severity of exposure bias and rollout stability. The following three schemes are evaluated: (i) \emph{full teacher forcing}, which conditions on ground-truth previous groups; (ii) \emph{scheduled sampling}, which mixes ground-truth and model-generated groups to partially match inference; and (iii) \emph{pure rollout training}, which always conditions on model predictions. Figure~\ref{fig:sample} shows that pure rollout quickly becomes unstable as the chain grows, indicating strong error accumulation when imperfect early groups are fed back as context, whereas teacher forcing remains stable but suffers from a train--test mismatch. Scheduled sampling provides a middle ground, substantially improving rollout stability by reducing this mismatch.

\begin{figure}[t]
  \centering
  \includegraphics[width=\columnwidth]{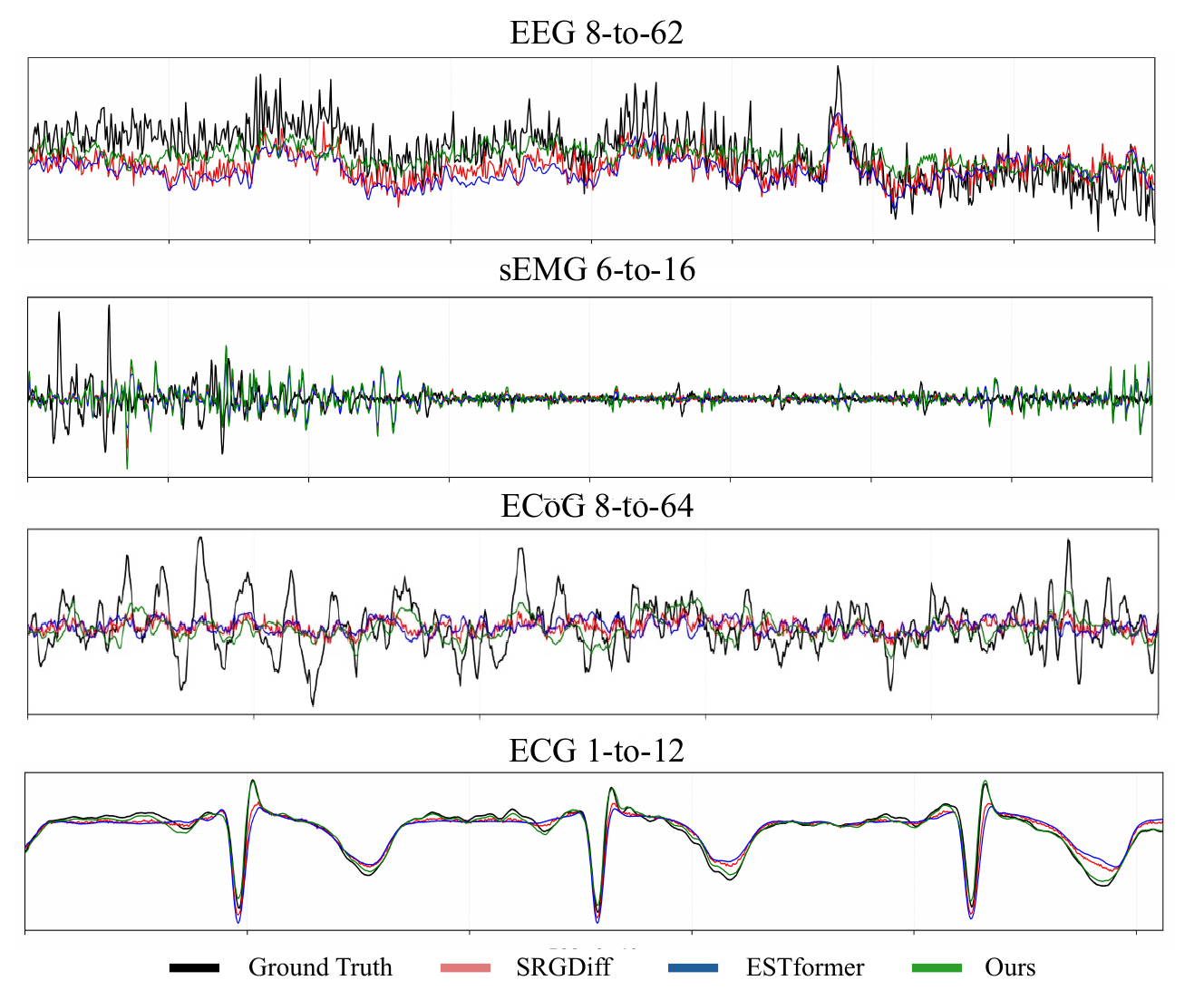}
  \vskip -0.1in
  \caption{Qualitative reconstruction comparisons across modalities and methods. }
  \vskip -0.1in
  \label{fig:tvis}
\end{figure}

\begin{figure*}[h]
  \centering
  \includegraphics[width=1.8\columnwidth]{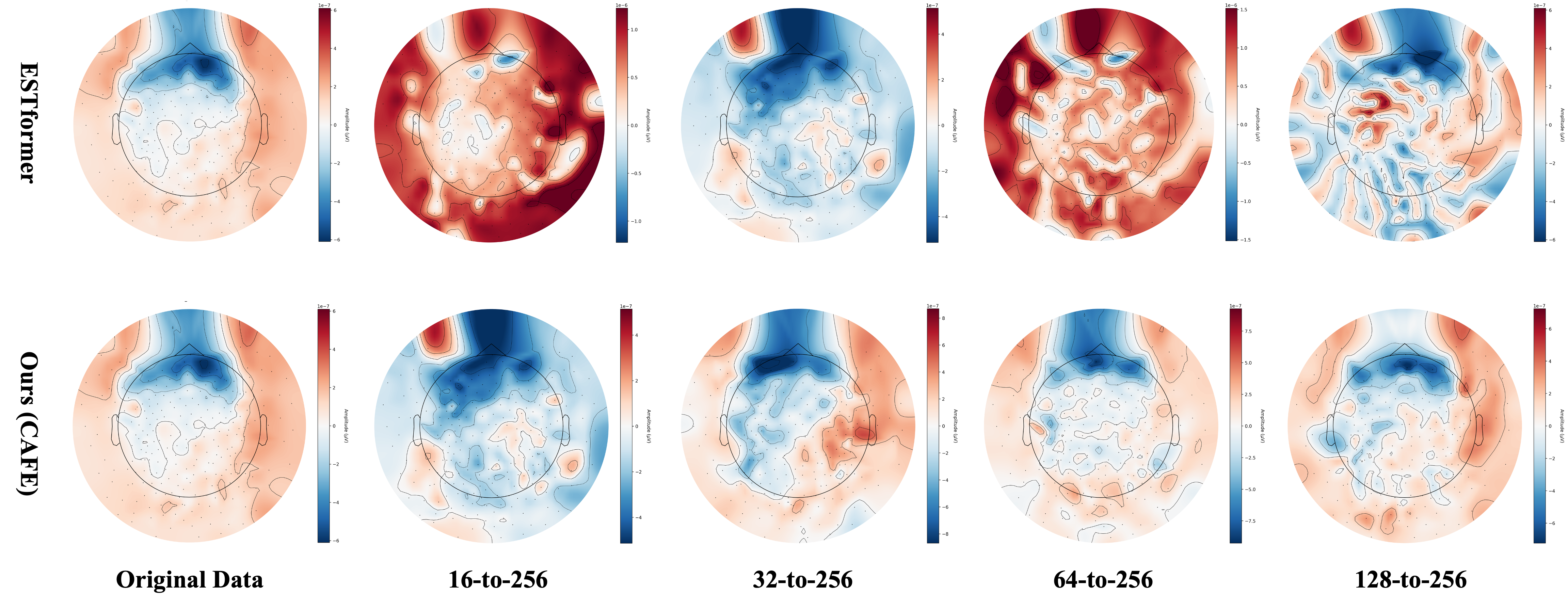}
  \vskip -0.1in
  \caption{Localize-MI spectral topomap visualization for two representative examples. Columns show the HD reference and reconstructions by ESTformer, SRGDiff, CGAN, and CAFE under the same SR setting.}
  \label{fig:mitopo}
\end{figure*}

\subsection{Baseline Comparisons}

\subsubsection{Baselines} 
CAFE is compared against representative and competitive baselines spanning (i) major modeling families for multichannel biosignal spatial super-resolution and (ii) widely used time-series reconstruction paradigms. 
For biosignal SR, the baselines include \textbf{ESTformer}~\citep{li2025estformer} (Transformer-based SR), \textbf{SRGDiff}~\citep{liu2025step} (diffusion-based SR), and \textbf{CGAN}~\citep{zhan2024conditional} (GAN-based SR). 
To connect with the broader time-series literature, results are also reported for a graph-based multivariate reconstruction method, \textbf{GRIN}~\citep{cinifilling}, and a strong time-series backbone, \textbf{TimeMixer++}~\citep{wang2025timemixer++}, instantiated as direct full-montage predictors under the same input/output protocol. 
All methods are trained and evaluated with identical preprocessing, data splits, and input length, and hyperparameters are tuned on the same validation set. Implementation details are provided in Appendix.

\subsubsection{Main Results}
Table~\ref{tab:cross_main} summarizes cross-subject channel SR across $6$ datasets and multiple upsampling factors. \textbf{CAFE} consistently achieves the strongest signal-level reconstruction, and its margin widens as the upsampling factor increases, indicating improved robustness when observations become sparser and the conditioning information is weaker. Among baselines, diffusion-based SRGDiff is the strongest competitor at high upsampling factors; however, its performance gap to CAFE increases as conditioning becomes sparser, indicating that stochastic denoising alone is insufficient to enforce topology-consistent dependency expansion. Table~\ref{tab:topomap} provides complementary frequency-domain evidence: CAFE yields substantially lower Spec-MAE across all scales (e.g., SEED $2\times$: $0.28$ vs.\ $0.376$--$0.511$; Localize-MI ($\times 10^{-7}$) $2\times$: $0.028$ vs.\ $0.108$--$0.145$), showing that the improvements extend beyond waveform fidelity to better preservation of spatial--spectral organization.
 CAFE exhibits substantially improved robustness under cross-layout and irregular electrode generalization, where the training and test sensor layouts differ, indicating that the proposed topology-aware rollout does not overfit a specific missing pattern. Robustness to noisy LD observations is further evaluated by injecting Gaussian noise into the LD anchors, and CAFE remains the most stable method as noise increases, with smaller degradation than competing baselines.

\subsubsection{Visualization Results}
We visualize frequency-domain topomaps on Localize-MI across methods (Figure~\ref{fig:mitopo}). CAFE better matches the HD reference in spatial localization and polarity, whereas ESTformer and CGAN more often exhibit shifted or over-smoothed patterns. SRGDiff is generally stable but can deviate in finer spatial organization. Complete topomap visualizations on SEED and Localize-MI are included in Figures~\ref{fig:mitopo}.
Moreover, qualitative time-domain trace comparisons across modalities are shown in Figure~\ref{fig:tvis}. CAFE more faithfully preserves peak timing and amplitude dynamics while reducing spurious oscillations, leading to smoother yet not over-damped reconstructions compared to other baselines. Additional examples are provided in Appendix.

\subsubsection{Downstream-level Comparison}
Downstream accuracy under different SR factors are summariezed in Appendix. These downstream tasks serve as functional probes, evaluating whether reconstructed signals preserve task-relevant discriminative structure beyond waveform-level fidelity. As the upsampling factor increases, accuracy decreases for all methods, yet the drop is consistently smaller for CAFE, suggesting improved robustness to sparse error impact via group-wise rollout. On high-density SEED, the advantage is most visible at larger factors: at $8\times$, CAFE reaches $0.58$ while the strongest baseline is $0.53$, indicating stronger retention of classification patterns under severe channel sparsity.

\subsubsection{Efficiency Comparison.}
Inference efficiency is compared in terms of trainable parameters, GFLOPs, and wall-clock latency for completing a full reconstruction.
CAFE is instantiated with the lightweight convolutional predictor, and the reported GFLOPs and latency are end-to-end values summed over all $G$ forward passes.
GFLOPs are computed on a $4$\,s input window under the same missingness setting as the main experiments. Latency (ms) is measured as the average end-to-end time per sample on the RTX $4090$ with fixed input length, averaged over $100$ runs.
As shown in Table~\ref{tab:comp_efficiency}, CAFE yields the lowest total compute (1.38 GFLOPs) and the fastest end-to-end latency (3.92 ms), while using substantially fewer trainable parameters (2.34M) than other baselines.

\begin{table}[h]
\centering
\caption{Computational cost and runtime of different SR models.}
\label{tab:comp_efficiency}
\small
\begin{tabular}{lccc}
\toprule
\textbf{Method} 
& \textbf{\#Params (M)} 
& \textbf{GFLOPs} 
& \textbf{Runtime (ms)} \\
\midrule
SRGDiff   
& 20.421     & 5.439    & 92.541 \\
ESTformer 
& 12.111 & 4.302 & 5.039 \\
CGAN      
& 39.309 & 6.093 & 134.106 \\
GRIN
& 0.079 & 14.803& 657.573 \\
TimeMixer++      
& 2.913 & 2.042 & 6.436 \\
CAFE   
& 2.342  & 1.380 & 3.922  \\
\bottomrule
\end{tabular}
\end{table}

\section{Conclusion}
In this work, we present a plug-and-play biosignal spatial super-resolution framework named CAFE. It imposes a group-wise, step-by-step generation constraint on existing backbones, replacing one-shot reconstruction with group-wise autoregressive rollout along spatial dimension. Using a shared predictor with masked canonical inputs and next-group supervision, CAFE consistently improves missing-channel reconstruction quality and robustness under severe channel sparsity across multiple modalities and datasets, while remaining computationally efficient with low GFLOPs and fewer trainable parameters than strong baselines.

\section*{Impact Statement}

This paper presents work whose goal is to advance the field of Machine
Learning. There are many potential societal consequences of our work, none
which we feel must be specifically highlighted here.

\bibliography{example_paper}
\bibliographystyle{icml2026}

\end{document}